\documentclass[journal=jacsat,manuscript=article]{achemso}

\usepackage{chemformula} 
\usepackage[T1]{fontenc} 
\usepackage{amsmath,amssymb,amsfonts,mathrsfs}
\usepackage{amsthm}
\usepackage{CJK}
\usepackage{bm}
\usepackage{lipsum}
\usepackage{babel}
\usepackage{graphicx}
\allowdisplaybreaks
\usepackage{makecell}
\usepackage{booktabs}
\usepackage{multirow}
\usepackage{gensymb}
\usepackage{rotating}
\usepackage{float}
\usepackage{braket}
\usepackage[breaklinks]{hyperref}
\usepackage{color}
\hypersetup{colorlinks=true, linkcolor=black, citecolor=black, filecolor=black, urlcolor=black}

\author{Jinyang Ni}
\email{jinyang.ni@ntu.edu.sg}
\affiliation{%
 Ministry of Education Key Laboratory for Nonequilibrium Synthesis and Modulation of Condensed Matter, Shaanxi Province Key Laboratory of Advanced Functional Materials and Mesoscopic Physics, School of Physics, Xi'an Jiaotong University, Xi'an 710049, China 
}%
\alsoaffiliation{%
 Division of Physics and Applied Physics, School of Physics and Mathematical Sciences, Nanyang Technological University, Singapore 637731, Singapore 
}%
\alsoaffiliation{These authors contributed equally to this work.}

\author{Zhenlong Zhang}
\affiliation{%
 Ministry of Education Key Laboratory for Nonequilibrium Synthesis and Modulation of Condensed Matter, Shaanxi Province Key Laboratory of Advanced Functional Materials and Mesoscopic Physics, School of Physics, Xi'an Jiaotong University, Xi'an 710049, China 
}%
\alsoaffiliation{These authors contributed equally to this work.}

\author{Jinlian Lu}
\affiliation{%
Department of Physics, Yancheng Institute of Technology, Yancheng, Jiangsu 224051, China 
}%
\author{Quanchao Du}
\affiliation{%
 Division of Physics and Applied Physics, School of Physics and Mathematical Sciences, Nanyang Technological University, Singapore 637731, Singapore 
}%

\author{Zhijun Jiang}
\email{zjjiang@xjtu.edu.cn}
\affiliation{%
 Ministry of Education Key Laboratory for Nonequilibrium Synthesis and Modulation of Condensed Matter, Shaanxi Province Key Laboratory of Advanced Functional Materials and Mesoscopic Physics, School of Physics, Xi'an Jiaotong University, Xi'an 710049, China 
}%
\alsoaffiliation{Key Laboratory of Computational Physical Sciences (Ministry of Education), Institute of Computational Physical Sciences, State Key Laboratory of Surface Physics and Department of Physics, Fudan University, Shanghai 200433, China}
\author{Laurent Bellaiche}
\affiliation{%
Smart Ferroic Materials Center, Physics Department and Institute for Nanoscience and Engineering, University of Arkansas, Fayetteville, Arkansas 72701, USA
}%
\alsoaffiliation{%
Department of Materials Science and Engineering, Tel Aviv University, Ramat Aviv, Tel Aviv 6997801, Israel
}%

\title{Nonvolatile Magnonics in Bilayer Magnetic Insulators}

\begin{document}
	\newpage
		
	\begin{center}
		\textbf{Abstract}
	\end{center}
	\baselineskip 22 pt
Nonvolatile control of spin order or spin excitations offers a promising avenue for advancing spintronics; however, practical implementation remains challenging. In this letter, we propose a general framework to realize electrical control of magnons in 2D magnetic insulators. We demonstrate that in bilayer ferromagnetic insulators with strong spin-layer coupling, electric field $E_{z}$ can effectively manipulate the spin exchange interactions between the layers, enabling nonvolatile control of the corresponding magnons. Notably, in this bilayer, $E_{z}$ can induce nonzero Berry curvature and orbital moments of magnons, the chirality of which are coupled to the direction of $E_{z}$. This coupling facilitates $E_{z}$
manipulate the corresponding magnon valley and orbital Hall currents. Furthermore, such bilayers can be easily engineered, as demonstrated by our density-functional-theory calculations on Janus bilayer $\mbox{Cr}$-based ferromagnets. Our work provides an important step toward realizing nonvolatile magnonics and paves a promising way for future magnetoelectric coupling devices.    
    ~~\\
	~~\\
	~~\\
	~~\\
	\textbf{Keywords:} Nonvolatile control, Magnons, Berry curvature, Orbital moments, Magnon Hall current

	\newpage

	
\textit{Introduction.} In ordered magnets, spins tend to align in a regular pattern due to strong exchange interactions\cite{neel1948proprietes, keffer1952theory, anderson1950antiferromagnetism}. When one or more spins are perturbed, this disturbance can propagate through the material as a wave, representing a collective excitation of the spins. The quanta of spin waves, known as magnons\cite{chumak2015magnon, lenk2011building}, play a fundamental role in understanding magnetic properties\cite{prabhakar2009spin, pirro2021advances}. Analogous to electric currents, magnon-based currents can be used to carry, transport and process information\cite{chumak2015magnon, onose_science_2010_329, katsura_prl_2010_104, matsumoto_PRB_2014_89, zyuzin2016magnon, cheng2016spin, li_PRL_2017_119, bao_nc_2018_9, mook_PRX_2021_11, chen_PRX_2018_8, yao_natphysics_2018_14, lee_prb_2018_97, zhang_PRL_2021_127, mcclarty2022topological}. In particular, magnons can possess nanometer-scale wavelengths\cite{chumak2014magnon, neusser2009magnonics}, operate in the terahertz (THz) frequency range\cite{li2020spin, zhang2024terahertz}, and enable the transfer of spin information over macroscopic distances\cite{chumak2014magnon, lebrun2018tunable}. These characteristics pave new avenues for wave-based computing technologies that are Joule-heat-free, addressing the fundamental limitations of modern electronics. 

In practical implementations, nonvolatile manipulation of magnons, particularly through electric fields, remains a significant challenge for various reasons\cite{bader2010spintronics, chumak2014magnon, jungwirth2016antiferromagnetic}. One of the key challenge is to establish an efficient coupling between magnons and electric field\cite{liu2021electric, parsonnet2022nonvolatile}. For instance, to control the magnon density, a substantial electric field induced modification of spin-wave dispersions is generally required\cite{rovillain2010electric, rana2019towards}. A possible route toward this goal involves the use of magnetoelectric materials capable of modulating spin exchange interactions via the electric
field. However, such materials have yet to be synthesized
experimentally\cite{eerenstein2006multiferroic, fiebig2016evolution, spaldin2010multiferroics, spaldin2019advances, dong2015multiferroic,dong2019magnetoelectricity}. Furthermore, the magnon quantity carrying spin information, such as spin or orbital angular momentum, are typically
decoupled from the electric field\cite{neumann2020orbital, fishman2022orbital, go2024magnon}. Therefore, developing a comprehensive strategy for an effective electric field control of magnons is highly desired.
\begin{figure}
    	\centering
    	\includegraphics[scale=0.55]{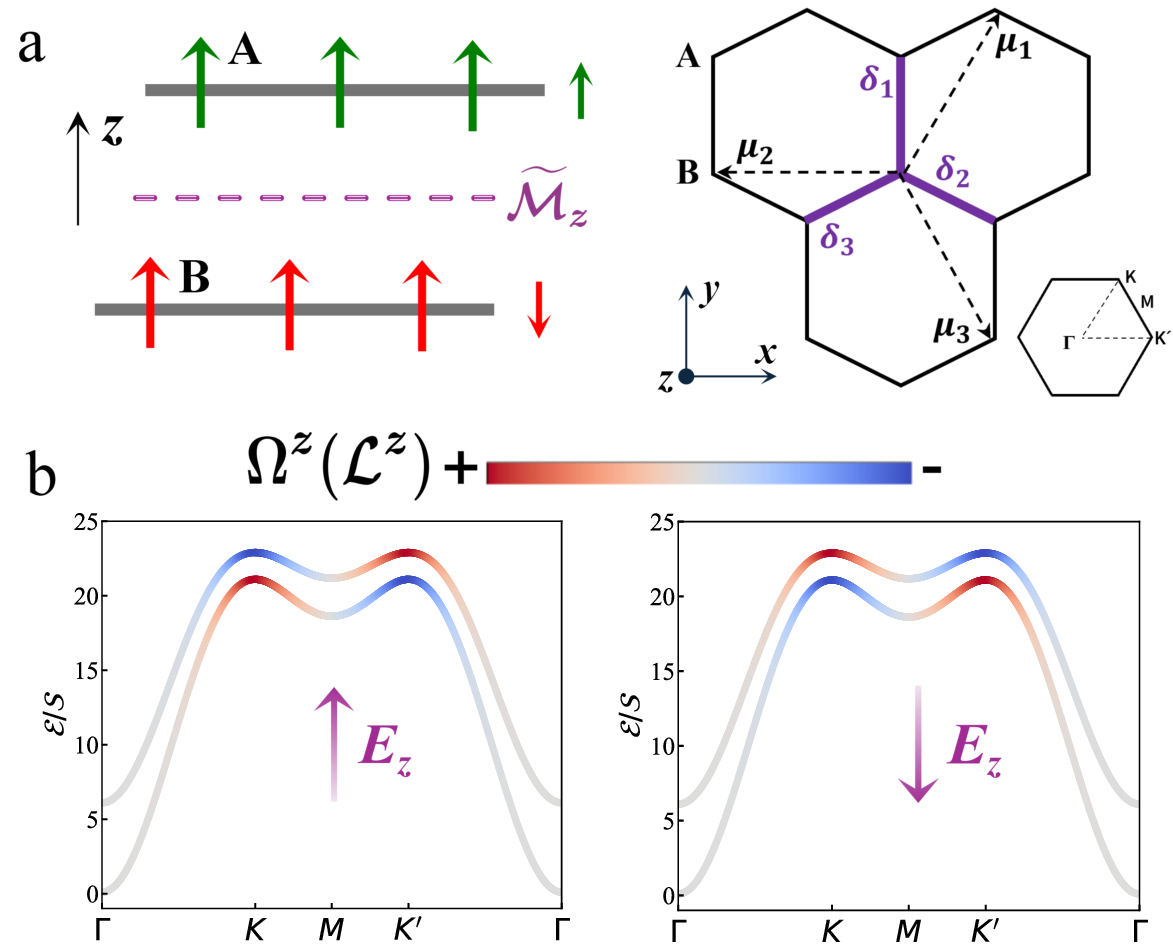}
    	\caption{(a)Left: side view of the of AB-stacked bilayer 
         triangular ferromagnet, where the polar vectors in top and bottom layers are opposite oriented. The $\widetilde{{\cal M}}_{z}$ combines both horizontal mirror symmetry and translation symmetry. Right: top view of such a bilayer ferromagnet. (b) The illustration of electric field control of the orbital moments and Berry curvature of magnons.
    \label{figure-1}}
\end{figure}

In this work, we present a general framework to realize electrical control of magnons in 2D magnetic insulators. We demonstrate that in bilayer ferromagnetic insulators, where each layer breaks the inversion symmetry and two layers connected by the mirror symmetry ${\cal M}_{z}$, strong spin-layer coupling can emerge. This spin-layer coupling enables the out-of-plane electric field $E_{z}$ to manipulate the spin exchange interactions between the layers, realizing a nonvolatile control of its magnons. By performing spin model analysis and density-functional-theory (DFT) calculations, we identify that $\mbox{Cr}$-based Janus bilayers exhibit strong spin-layer coupling with respect to $E_{z}$. Since $E_{z}$ breaks the inversion symmetry, it can induce and manipulate the magnon Berry curvature as well as the magnon orbital moments. Additionally, $E_{z}$ can also generate orbital Berry curvature of magnons, which contribute to a significant magnon orbital Hall current. This mechanism enables ultrafast control of magnon Hall transport via an electric field.

\textit{General theory.} We firstly consider the AB-stacked
bilayer magnet, which consist of two triangular ferromagnets layers coupled through ferromagnetic interactions. As schematically illustrated
in Figure\,\ref{figure-1}a, this bilayer triangular ferromagnet is analogous
to a monolayer honeycomb ferromagnet. In this analogy, the nearest-neighbor
(NN) intralayer spin exchange within the upper or bottom layer corresponds
to the next-nearest-neighbor (2NN) spin exchange along the A-A or
B-B paths in the honeycomb lattice. Similarly, the NN interlayer spin
exchange corresponds to the NN intralayer spin exchange interaction
of a monolayer honeycomb lattice. Compared with the conventional bilayer ferromagnets, our bilayer model exhibits two key distinctions in symmetry: (i) the inversion symmetry ${\cal P}$ within each layer is broken, and (ii) the effective horizontal mirror reflection $\widetilde{{\cal M}}_{z}$ or inversion symmetry $\cal P$ of this bilayer is preserved. Protected by these symmetry operations, the polar vectors of the top and bottom layers are oppositely oriented, as shown in Figure\,\ref{figure-1}a. This feature gives rise to a robust magnetoelectric effect\cite{fiebig2005revival}, where the $E_{z}$ not only induces the net polarization but also modifies the spin polarization of this bilayer. Correspondingly, the spin associated quantity of wave functions within each layer can be tuned by $E_{z}$, thus enabling spin-layer coupling. This coupling, in turn, allows the spin exchange interactions to be controlled via the $E_{z}$.

Following the above symmetry requirement, the minimum spin model of
this bilayer ferromagnet can write as 
\begin{equation}\label{eq1}
\begin{split}{\cal \hat{H}}=  {\cal J}_{ab}\sum_{\langle i,j\rangle}{\cal S}_{i}\cdot{\cal S}_{j}+{\cal J}_{a}\sum_{\langle i,j\rangle}{\cal S}_{i}\cdot{\cal S}_{j}+{\cal J}_{b}\sum_{\langle i,j\rangle}{\cal S}_{i}\cdot{\cal S}_{j} +{\cal K}_{a}\sum_{i}{\cal S}_{i,z}^{2}+{\cal K}_{b}\sum_{i}{\cal S}_{i,z}^{2},
\end{split}
\end{equation}
where ${\cal J}_{ab}$ represents NN interlayer spin exchange interaction,
${\cal J}_{a(b)}$ denotes NN intralayer spin exchange interactions
in the top (bottom) layer, and ${\cal K}_{a(b)}$ refers to the easy-axis
single-ion-anisotropy (SIA) at top(bottom) layer. The linear spin
wave model in Eq.\,(\ref{eq1}) can be solved by employing Holstein-Primakoff
(HP) transformation\citep{holstein1940field}, ${\cal S}_{i,\alpha}^{z}=S-\hat{a}_{i,\alpha}^{\dagger}\hat{a}_{i,\alpha},{\cal S}_{i}^{+}\approx\sqrt{2S}\hat{a}_{i,\alpha}$
and ${\cal S}_{i,\alpha}^{-}\approx\sqrt{2S}\hat{a}_{i,\alpha}^{\dagger}$
where $\hat{a}_{i,\alpha}^{\dagger}$ creates a magnon on $\alpha$
sublattice $(\mbox{A},\mbox{B})$ in the $i$-th unit cell. Upon Fourier
transformation, the Hamiltonian can be expressed in the basis
$\psi_{k}^{\dagger}\equiv(\hat{a}_{k}^{\dagger},\hat{b}_{k}^{\dagger})$
as ${\cal {\hat{H}}}=\sum_{k}\psi_{k}^{\dagger}{\cal \hat{H}}_{k}\psi_{k}$.
Neglecting the zero-point energy, ${\cal \hat{H}}_{k}$ reads as,
\begin{equation}\label{eq2}
{\cal \hat{H}}_{k}/S =-3{\cal J}_{ab}+\left(\begin{matrix}{\cal J}_{a}f_{k}-2{\cal K}_{a} & {\cal J}_{ab}\gamma_{k}\\
{\cal J}_{ab}\gamma_{k}^{\dagger} & {\cal J}_{b}f_{k}-2{\cal K}_{b}
\end{matrix}\right),
\end{equation}
where $\gamma_{k}=\sum_{\delta}e^{-i\textbf{\textit{k}}\cdot\boldsymbol{\delta_{i}}}$ and $f_{k}=\sum_{i\in odd}2\mbox{cos}(\textbf{\textit{k}}\cdot\boldsymbol{\mu}_{i})-6$
with $\boldsymbol{\delta}_{i}$ and $\boldsymbol{\mu}_{i}$ being
NN and 2NN linking vectors of honeycomb lattice as shown in Figure\,\ref{figure-1}a.

For simplicity, Eq.\,(\ref{eq2}) can be parameterized as $h_{0}=\frac{1}{2}({\cal J}_{a}+{\cal J}_{b})f_{k}-({\cal K}_{a}+{\cal K}_{b})$,
$h_{x}={\cal J}_{ab}\mbox{Re}\gamma_{k}$, $h_{y}={\cal J}_{ab}\mbox{Im}\gamma_{k}$
and $h_{z}=\frac{1}{2}({\cal J}_{a}-{\cal J}_{b})f_{k}-({\cal K}_{a}-{\cal K}_{b})$. Eq.\,(\ref{eq2}) can thus be expressed in terms of Pauli-matrices
$\boldsymbol{\sigma}=\left(\sigma_{x},\sigma_{y},\sigma_{z}\right)$, given as
\begin{equation}\label{eq3}
{\cal \hat{H}}_{k}/S=h_{0}I+\boldsymbol{h}(\boldsymbol{k})\cdot\boldsymbol{\sigma},
\end{equation}
where $\boldsymbol{h}(\boldsymbol{k})=\left(h_{x},h_{y},h_{z}\right)$
= $|h|\left(\mbox{sin}\theta\mbox{cos}\phi,\mbox{sin}\theta\mbox{sin}\phi,\mbox{cos}\theta\right)$.
The corresponding eigenvalues and eigenvectors are given by 
\begin{equation}\label{eq4}
{\epsilon}_{\pm}/S=h_{0}\pm|\boldsymbol{h}(\boldsymbol{k})|,\quad\Psi_{\pm}=\begin{pmatrix}\sqrt{1\pm\mbox{cos}\theta}\\
\pm e^{-i\phi}\sqrt{1\mp\mbox{cos}\theta}
\end{pmatrix}.
\end{equation}
Clearly, the emergence of $h_{z}$ breaks the sublattice symmetry of honeycomb lattice and open a gap of magnon bands
at $\mbox{K}$-points. Under this condition, the time reversal symmetry ($\cal T$) is preserved and the the inversion symmetry ${\cal P}$ is broken\cite{mook_PRX_2021_11}, resulting in a nonzero
Berry curvature. In 2D cases, the Berry curvature only have $z$-component, which can be expressed in term of the $\boldsymbol{h}(\boldsymbol{k})$
vectors, 
\begin{equation}\label{eq5}
\boldsymbol{\Omega}_{\pm}^{z}(\boldsymbol{k}) =-i\langle\boldsymbol{\nabla}\Psi_{\pm}|\times|\boldsymbol{\nabla}\Psi_{\pm}\rangle =\mp\frac{h_{z}}{2}\left(\boldsymbol{\nabla}h_{x}\times\boldsymbol{\nabla}h_{y}\right).
\end{equation}
Consequently, in this bilayer, the nonzero Berry curvature depend on the values of ${\cal J}_{a} - {\cal J}_{b}$ and ${\cal K}_{a} - {\cal K}_{b}$, indicating the chirality of magnon Berry curvature can be controlled by ${E}_{z}$. Notably, the magnon Berry curvature is absent in the monolayer due to the presence of only a single magnon branch.
	
    \begin{figure}
    	\centering
    	\includegraphics[scale=0.45]{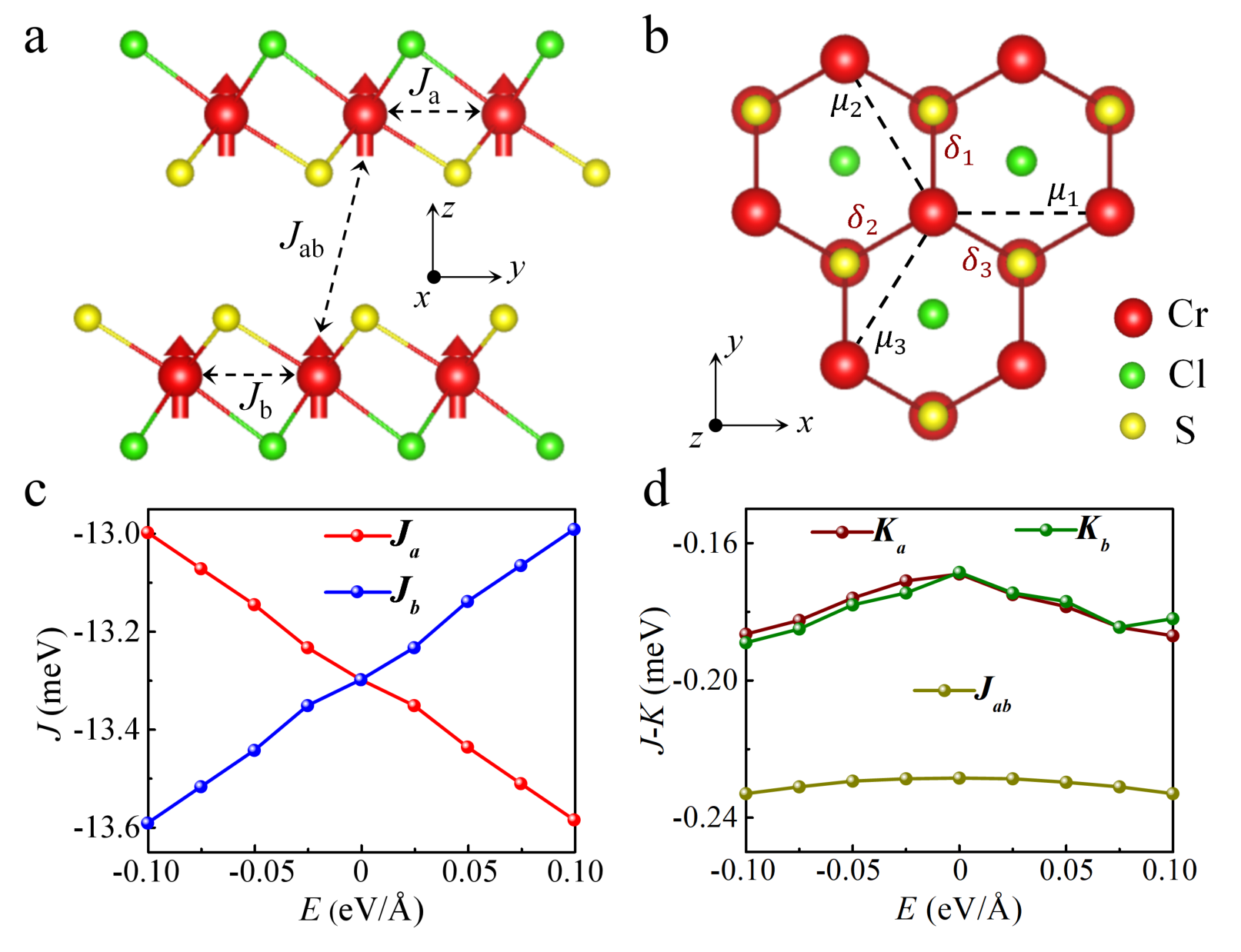}
    	\caption{(a) Side view of the geometrical structure of AB-stacked bilayer $\mbox{CrSCl}$. (b) Top view of the bilayer $\mbox{CrSCl}$. (c)-(d) The dependence of intralayer spin exchange interactions ${\cal J}_{a}({\cal J}_{b})$, interlayer spin exchange interaction ${\cal J}_{ab}$ and SIA ${\cal K}_{a}({\cal K}_{b})$ on $E_{z}$ in bilayer $\mbox{CrSCl}$.
    		\label{figure-2}}
    \end{figure}

In addition to the Berry curvature, the nonzero $h_{z}$ can also
induce orbital moments of magnons\cite{neumann2020orbital, fishman2022orbital, go2024magnon}.
To study this orbital quantity, we invoke the semiclassical formulation
of the wave packet dynamics of Bloch magnons\cite{chang1996berry, xiao2010berry},
which is given by 
\begin{equation}\label{eq6}
\boldsymbol{{\cal L}}_{m,\pm}^{z}(\boldsymbol{k})=-\frac{i}{2\hbar}\langle\boldsymbol{\nabla}\Psi_{\pm}|\times(\hat{{\cal H}}_{k}-\bar{\epsilon}_{k})|\boldsymbol{\nabla}\Psi_{\pm}\rangle.
\end{equation}
Compared to the Berry curvature of magnons, the orbital moments
also depend on the $h_{z}$; however, the chirality of orbital moments is identical for the majority and minority bands. Note that the orbital angular momentum (${\cal L}^{z}$) and orbital moments
(${\cal L}_{m}^{z}$) of magnons differ only by a factor related to mass term. Therefore, Berry curvature and orbital moments of magnons can be induced and manipulated by electric field in the bilayer
ferromagnetic insulators with strong spin-layer coupling.
	
\textit{Layer spin coupling in bilayer $\mbox{Cr}XY$.} Having established the concept of electrically switchable magnons in bilayer ferromagnets based on the effective model, we next illustrate this with a concrete example of Janus bilayer $1\mbox{T}$-$\mbox{Cr}XY$ ($X=\mbox{S},\mbox{Se};Y=\mbox{Cl},\mbox{Br}$). Janus monolayer $1\mbox{T}$-$\mbox{Cr}XY$ is derived from the well-known monolayer $\mbox{Cr}$-based dichalcogenide $1\mbox{T}$-$\mbox{CrTe}_{2}$,
where the chalcogen layer substituted by halides, leading to in a transition from the metallic to insulator states\citep{zhang2021room, meng2021anomalous, xian2022spin, hou2022multifunctional, xiao2020two}. In monolayer $\mbox{Cr}XY$, the $\mbox{Cr}^{3+}$ ions with $S=3/2$ form a triangular lattice, where the inversion symmetry ${\cal P}$ is broken. Building upon this monolayer, bilayer $\mbox{Cr}XY$ adopts an AB-stacking configuration
with $Y$-$X$-$X$-$Y$ order, preserving the ${\cal P}$, as illustrated in Figure\,\ref{figure-2}a. The structure stability of bilayer Janus
$1\mbox{T}$-$\mbox{Cr}XY$ is confirmed through the phonon dispersion
calculation and molecular dynamics simulation, with details shown
in Supporting Information (SI\cite{Supplemental_Materials}). By employing energy mapping method\cite{xiang2011predicting,xiang2013magnetic}, we calculate the interlayer and intralayer spin exchange interactions of bilayer $\mbox{Cr}XY$, both of which are ferromagnetic, as shown in Table \ref{tab_1}, confirming its the ferromagnetic ground state.
It is noteworthy that the magnetic anisotropy of bilayer $\mbox{CrS}Y$
and $\mbox{CrSe}Y$ differs; the former exhibits easy-axis, while
the latter shows easy plane, as indicated by the sign of ${\cal K}$. As shown in Figure\,S7\cite{Supplemental_Materials}, the calculated orbital-resolved magnetic anisotropy energy (MAE) reveals that the MAE of $\mbox{Cr}XY$ mainly depends on the hybridizations of 3$d$-orbitals of $\mbox{Cr}^{3+}$ between $p$-orbitals of $X^{2-}$ and $Y^{-}$. For $\mbox{CrS}Y$, the dominant contributions to the easy-axis MAE originate from $\mbox{Cr}^{3+}$ and $Y^{-}$ ions, resulting in easy-axis magnetic anisotropy. In contrast, significant easy-plane contributions from $\mbox{Se}^{2-}$ drive the easy-plane magnetic anisotropy of $\mbox{CrSe}Y$. Unless otherwise
specified, the following discussion primarily focuses on the bilayer $\mbox{CrSCl}$ for simplicity.

\linespread{1.3}
    \begin{table}[!h]
    	\caption{Calculated spin exchange parameters (with units of meV)
of Janus bilayer $\mbox{Cr}{XY}$ without $E_{z}$.}
    	\begin{tabular}{cccc}
    	   \hline
    	   $Mat.$   &   ${\cal J}_{ab}$    &  ${\cal J}_{a}$ (${\cal J}_{b}$) &${\cal K}_{a}$(${\cal K}_{b}$) \\
    	   \hline
    	   $\mbox{CrSCl}$ &  -0.23  & -13.31  &-0.17 \\
    	 $\mbox{CrSBr}$ & -0.21  & -14.48  &-0.23 \\
    	    {$\mbox{CrSeCl}$} & -0.24  & -11.77  & 0.21 \\
    	    $\mbox{CrSeBr}$ & -0.11  & -14.49 &0.11 \\
    	   \hline
    	\end{tabular}
    	\label{tab_1}
    \end{table}

When $E_{z}=0$, the $\cal P$ ensures that the localized spin moments of $\mbox{Cr}^{3+}$ ions within each layer of bilayer $\mbox{CrSCl}$ are identical. Consequently, the NN intralayer spin exchange interactions and SIA between the top and bottom layers are identical, i.e., ${\cal J}_{a} = {\cal J}_{b}$ and ${\cal K}_{a} = {\cal K}_{b}$. Upon applying $E_{z}$, ${\cal P}$
is naturally broken, disrupting this balance. Specially, for $E_{z}=0.1\ \mbox{eV/\AA}$,
the difference in localized spin moments of $\mbox{Cr}^{3+}$ ions between two layers reaches approximately $0.005\ \mu_{B}$.
This imbalance is notably reflected in the remarkable changes to the spin exchange
parameters between the two layers, as illustrated in Figure\,\ref{figure-2}c, where
${\cal J}_{b}-{\cal J}_{a}=0.6\ \mbox{meV}$ with $E_{z}=0.1\ \mbox{eV/\AA}$. As
the strength of $E_{z}$ increases, ${\cal J}_{a}-{\cal J}_{b}$
exhibits a linear dependence on $E_{z}$, described by ${\cal J}_{a}-{\cal J}_{b}=-6E_{z}$, indicating that electric field can reverse the sign of ${\cal J}_{a}-{\cal J}_{b}$. In contrast to the Heisenberg exchange interactions, the change in ${\cal K}_{a}-{\cal K}_{b}$ with respect to $E_{z}$ is slower due to the weak SOC of bilayer $\mbox{CrSCl}$, so we can assume ${\cal K}_{a}={\cal K}_{b}$
in the following discussion. Similarly, the Dzyaloshinskii-Moriya interaction of bilayer $\mbox{CrSCl}$ is relatively weak\cite{hou2022multifunctional}, and its response to $E_{z}$ is negligible. In addition, the NN intralayer coupling, ${\cal J}_{ab}$, can be enhanced by $E_{z}$, reaching $-0.25\ \mbox{meV}$ when $E_{z}$ = $0.2\ \mbox{eV/\AA}$. According to Eqs.\,(\ref{eq5})-(\ref{eq6}), a significant ${\cal J}_{a}-{\cal J}_{b}$ can give rise to the nonzero Berry curvature and orbital moments of magnons, with their chirality being coupled to the ${E_{z}}$. 

\begin{figure}
		\centering
		\includegraphics[scale=0.35]{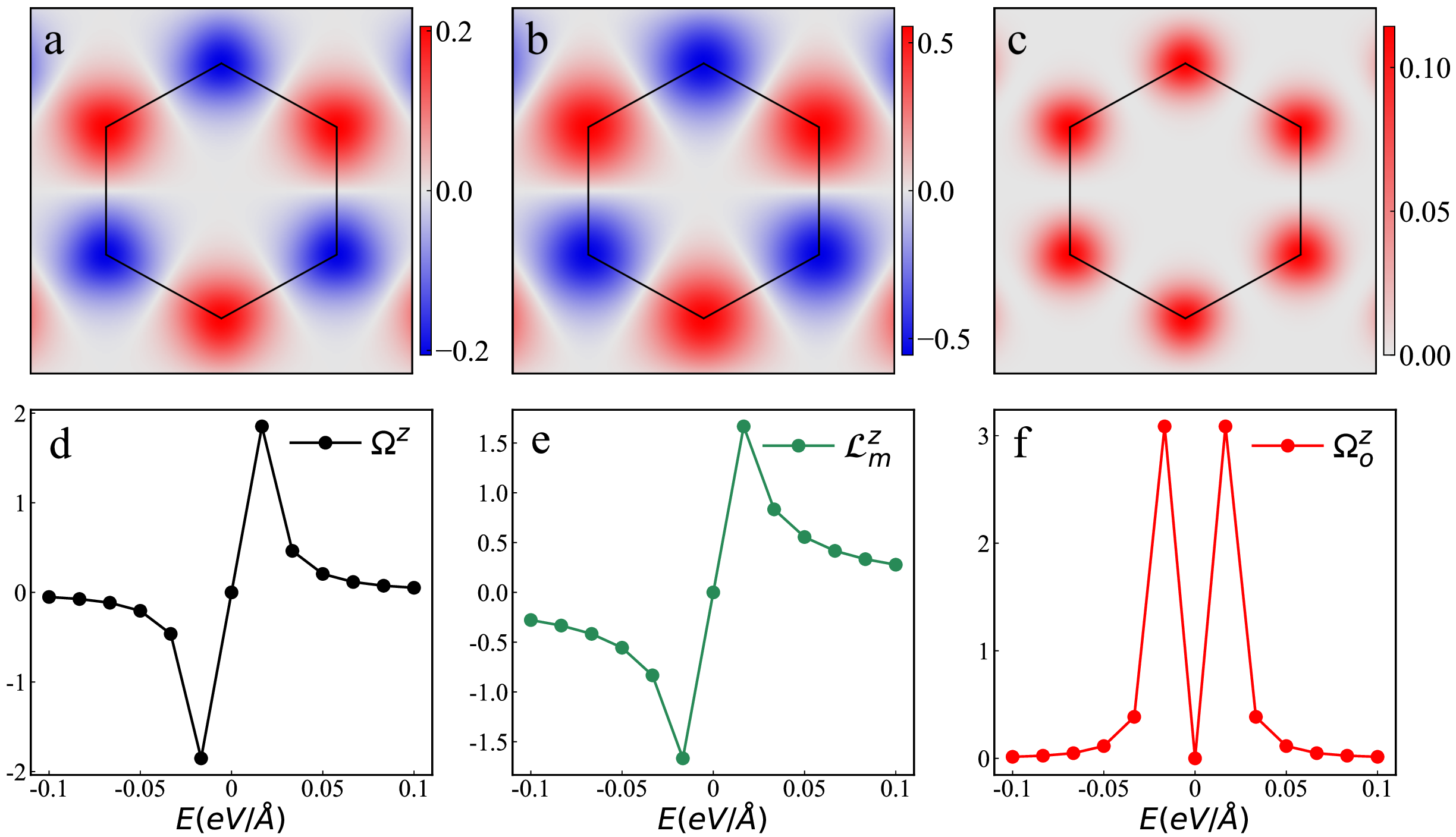}
		\caption{(a)-(c) Berry curvature ($\Omega^{z}$), orbital moments (${\cal L}^{z}_{m}$), and orbital Berry curvature ($\Omega^{z}_{o}$) of minor magnon band in bilayer $\mbox{CrSCl}$ with $E_{z} = 0.05\ \mbox{eV/\AA}$. (d)-(f) The dependence of $\Omega^{z}$, ${\cal L}^{z}_{m}$, and $\Omega^{z}_{o}$ at $\mbox{K}$-points on $E_{z}$.
			\label{figure-3}}
\end{figure}

\begin{figure}
		\centering
		\includegraphics[scale=0.4]{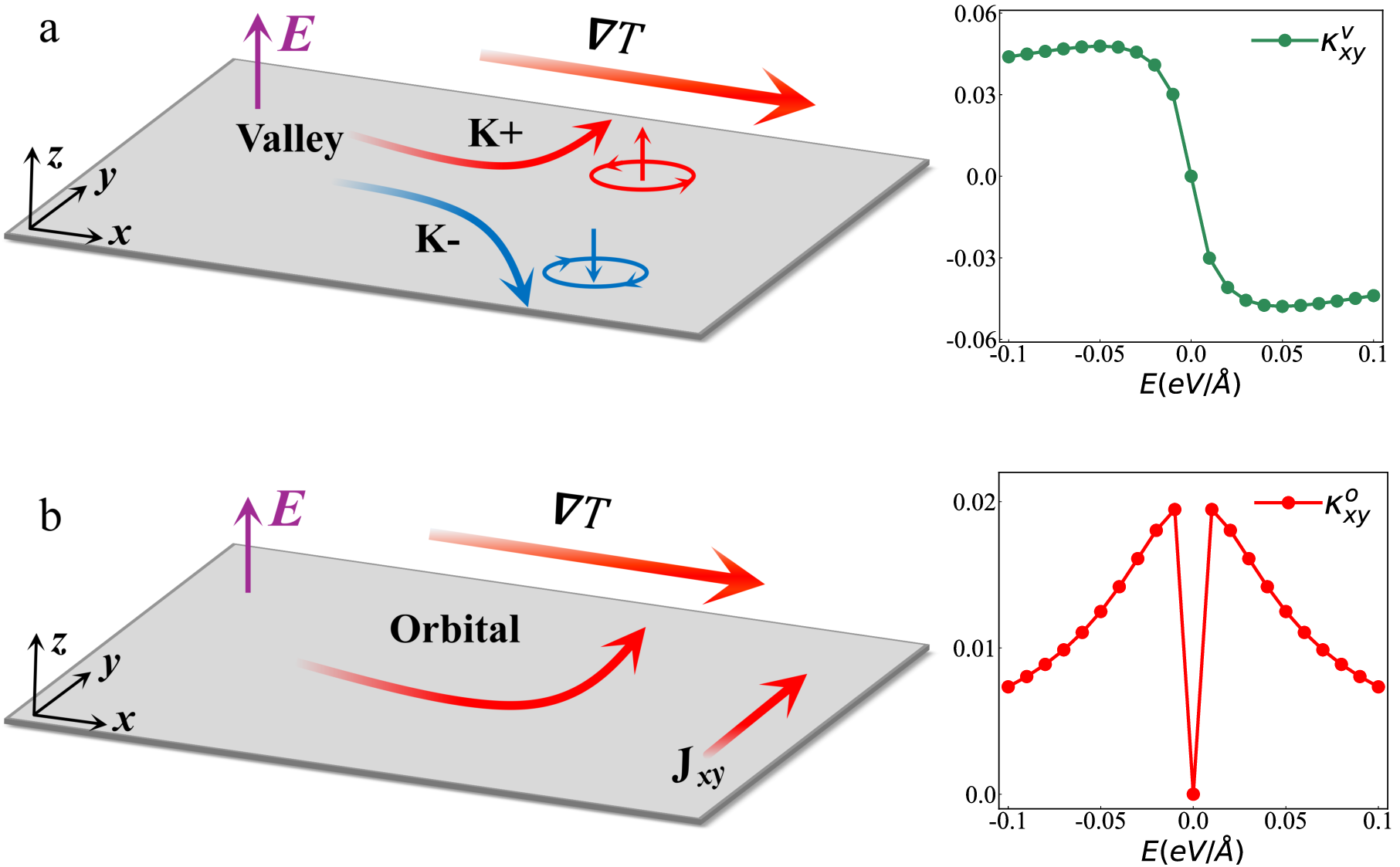}
		\caption{(a)The left panel illustrates the magnon valley Hall effect.  The right panel shows the dependence of the magnon valley conductivity ${\cal \kappa}_{xy}^{v}$ on $E_{z}$. (b) The left panel illustrates the magnon orbital Hall effect. The right panel shows the dependence of the orbital Hall conductivity ${\cal \kappa}_{xy}^{o}$ on the $E_{z}$. Both ${\cal \kappa}_{xy}^{v}$ and ${\cal \kappa}_{xy}^{o}$ are expressed in units of $k_{B}^{2}/{\hbar}$.
			\label{figure-4}}
\end{figure}

\textit{Nonvolatile control of magnon transport.} In Figure\,\ref{figure-3}a,
we observe that the Berry curvature of magnons at $E_{z}=0.05\,\mbox{eV/Å}$ is primarily
concentrated at the ${\mbox{K}}$-points and satisfy $\boldsymbol{\Omega}(\boldsymbol{k})=-\boldsymbol{\Omega}(-\boldsymbol{k})$,
indicating the inversion symmetry ${\cal P}$ of magnons is broken.
In this case, the net magnon Hall current, the first order thermal Hall response under the temperature gradient, remains zero due to the opposite contributions of magnon Berry curvature at $\mbox{K}$ and $\mbox{K}^{\prime}$-points. However, the opposite orbit pseudo-magnetic field at $\mbox{K}$ and $\mbox{K}^{\prime}$-points give rise to the magnon valley current, similar to the gapped graphene\cite{xiao2007valley, bhowal2021orbital}. To distinguish
the difference between $\mbox{K}$ and $\mbox{K}^{\prime}$-points \cite{zhai2020topological},
we define the magnon valley conductivity ${\cal \kappa}^{v}_{xy}$ as 
\begin{equation}\label{eq7}
{\cal \kappa}_{xy}^{v}=\frac{k_{B}^{2}T}{\hbar V}\sum_{n,k}c_{2}(\rho)\left[\Omega_{n,k}(\mbox{K})-\Omega_{n,k}(\mbox{K}^{\prime})\right],
\end{equation}
where $V$ is the volume, $\varepsilon_{abc}$ is the Levi-Civita
symbol with $abc=xyz$ and $c_{2}(\rho)=\int_{0}^{\rho}[\mbox{log}(1+\rho^{-1})]^{2}d\rho$ with $\rho$ being Boltzmann distributions for bosons. Figure\,\ref{figure-4}a illustrates the nonlinear dependence of ${\cal \kappa}_{xy}^{v}$ on $E_{z}$ at a fixed temperature. Specifically, ${\cal \kappa}_{xy}^{v}$ starts from zero, reaches maximum at $E_{z}=0.05\ \mbox{eV/\AA}$ (${\cal J}_{a}-{\cal J}_{b}=-0.3\ \mbox{meV}$), then decreases. Since the $E_{z}$ can reverse the sign of ${\cal J}_{a}-{\cal J}_{b}$, the chirality of Berry curvature at $\mbox{K}$-points and the sign of ${\cal \kappa}_{xy}^{v}$ can be electrically controlled.

As shown in Eq.\,(\ref{eq6}), in addition to the Berry curvature, the nonzero $h_{z}\approx({\cal J}_{a}-{\cal J}_{b})f_{k}$ can induce the magnon orbital moments, indicating that $E_{z}$ can effectively manipulate the magnon orbital quantity. Figure\,\ref{figure-3}b illustrates the orbital moments of magnons at $E_{z}=0.05\ \mbox{eV/\AA}$, indicating that orbital moments primarily reside at $\mbox{K}$-points and exhibit opposite signs at different $\mbox{K}$-points, similar to behavior of the Berry curvature. Because ${\cal J}_{a}-{\cal J}_{b}$ is governed by $E_{z}$, the orbital moments or orbital angular momentum of magnons can be switched electrically, as evidenced in Figure\,\ref{figure-3}e.

The emergence of magnon orbital moments gives rise to the anomalous
orbital associated transport phenomena\cite{go2024magnon, bhowal2021orbital}.
As shown in the SI\cite{Supplemental_Materials}, we derived the analytical expression for magnons
orbital Berry curvature, $\Omega_{o}^{z}$, defined as follows 
\begin{equation}\label{eq8}
\boldsymbol{\Omega}_{o}^{z}(\boldsymbol{k})=\boldsymbol{\Omega}^{z}(\boldsymbol{k})\boldsymbol{{\cal L}}_{m}^{z}(\boldsymbol{k}).
\end{equation}
Obviously, the orbital Berry curvature is the product of the Berry
curvature and the orbital moments of magnons\cite{go2024magnon, bhowal2021orbital}.
As both the Berry curvature and orbital moments are odd with respect
to ${\boldsymbol{k}}$, their product, orbital Berry curvature, is
even, $\boldsymbol{\Omega}_{o}^{z}(\boldsymbol{k})=\boldsymbol{\Omega}_{o}^{z}(-\boldsymbol{k})$,
breaking the effective time-reversal symmetry in orbital space. This
characteristic enables the emergence of the net magnon Hall current,
in contrast to the contribution from the Berry curvature alone, as
shown in Figure\,\ref{figure-4}. Thus, analogous to the magnon Hall effect,
the magnon orbital Hall conductivity is determined by integrating the product of $c_{2}(p)$ and $\boldsymbol{\Omega}^{z}_{o}$ over the first Brillouin zone, expressed as follows 
\begin{equation}\label{eq9}
{\cal \kappa}_{xy}^{o}=\frac{k_{B}^{2}T}{\hbar V}\sum_{n,k}c_{2}(\rho)\Omega_{n,o}^{z}(k). 
\end{equation}
Figure\,\ref{figure-4}b shows that the dependence of ${\cal \kappa}_{xy}^{o}$ on $E_{z}$ is even, which can be attributed to the even relationship between the orbital Berry curvature and $E_{z}$. Notably, ${\cal \kappa}_{xy}^{o}$ decreases with increasing $E_{z}$ and diverges as $E_{z}$ approaches zero. This divergence
arises because the Berry curvature and orbital moments
of each band vanish across most of the momentum space, except at
the $\mbox{K}$-points, when $E_{z}$ is small. As a result, the orbital Berry curvature diverges as $E_{z}$ approaches zero as well as the the orbital Hall conductivity.

\textit{Bilayer honeycomb ferromagnets.} The picture of electrical control of magnons can also be applied to Janus bilayer honeycomb ferromagnets, such as $\mbox{Cr}_{2}\mbox{Br}_{3}\mbox{Cl}_{3}$\cite{huang2017layer, xu2020topological}. The ground state of Janus bilayer $\mbox{Cr}_{2}\mbox{Br}_{3}\mbox{Cl}_{3}$ adopts AB-stacked configuration, where the intralayer and interlayer
coupling are both ferromagnetic, as confirmed by our DFT calculations presented in SI\cite{Supplemental_Materials}. When $E_{z}=0$, the ${\cal PT}$ symmetry of its magnons is preserved, and the corresponding magnon bands exhibits three-fold degeneracy at $\mbox{K}$-points. When $E_{z}$ is applied, the local spin moments of $\mbox{Cr}^{3+}$ ions will differ between layers, similar to the Janus bilayer $\mbox{Cr}XY$. This spin-layer
coupling modifies the spin exchange interactions and SIA within each layer, breaking the inversion symmetry ${\cal P}$ of magnons. Consequently, this can lift the threefold degeneracy in the magnon bands at $\mbox{K}$-points, generating nonzero Berry curvature and orbital moments, which in turn induce magnon valley and orbital Hall currents. Compared with bilayer $\mbox{Cr}XY$, the sublattice symmetry of honeycomb lattice within each layer in bilayer $\mbox{Cr}_{2}\mbox{Br}_{3}\mbox{Cl}_{3}$ remains preserved even when $E_{z}\neq0$. Consequently, the
Berry curvature and orbital moments of corresponding magnon bands are even with respect to the $E_{z}$. 

In addition to the Janus bilayer, bilayer $\mbox{2H}$-type magnets, such as bilayer $\mbox{2H}$-$M\mbox{Se}_{2}$ with ($ M = \mbox{V, Cr, Mn}$),
also satisfy the symmetry requirement outlined above\cite{zhang2021room, bonilla2018strong,yu2019chemically, wang2021ferromagnetism, eren2019defect,zhang2017van,shen2024electrical, duvjir2018emergence}.
The numerous candidate materials provide strong evidence for the applicability
of our theory and offer a broader range of options for experimental
realization.

\textit{Summary and discussion.}  In summary, we propose a general framework for achieving electrical control of magnons in bilayer ferromagnetic insulators with strong spin-layer coupling. We demonstrate that the strong spin-layer coupling allows the electric field to manipulate the spin exchange interactions between layers, as evidenced by our DFT calculations on Janus bilayer $\mbox{Cr}$-based ferromagnetic insulators. This
capability facilitates the electrical control of strength and chirality of magnons, encompassing their Berry curvature, orbital moments, and orbital Berry curvature. Therefore, the magnon valley Hall and  orbital Hall transport can be effectively tuned by applying an electric field. The broad applicability of our findings, coupled with the diverse array of material candidates identified, lays a solid foundation for the development of high-speed, compact spintronic and magnonic devices.
    
	~~\\
	~~\\
	\textbf{ASSOCIATED CONTENT}\\
	\textbf{Supporting Information}\\
    In this Supporting Information, we provide: I. the details of DFT calculations and additional calculations of Janus bilayers $\mbox{CrSCl}$ and $\mbox{Cr}_{2}\mbox{Br}_{3}\mbox{Cl}_{3}$; II. the magnetic anisotropy of bilayer $\mbox{Cr}XY$. III. the details of the orbital Berry curvature of magnons; and IV. the magnon Hamiltonian
 of bilayer $\mbox{Cr}_{2}\mbox{Br}_{3}\mbox{Cl}_{3}$.

    ~~~\\
	~~\\
	\textbf{AUTHOR INFORMATION}\\
  Jinyang Ni and Zhenlong Zhang equally contributed to this work.\\
	\textbf{Corresponding Authors}\\
	$^*$E-mail: jinyang.ni@ntu.edu.sg (J. N.)\\
    $^*$E-mail: zjjiang@xjtu.edu.cn (Z. J.)\\
      
	\textbf{Notes}\\
	The authors declare no competing financial interest. 
	
	~~~\\
	~~~\\
	\textbf{ACKNOWLEDGMENTS}\\

The authors thank Prof. Yuanjun Jin for helpful discussions. This work is supported by the National Natural Science Foundation of China (Grant No. 12374092), Natural Science Basic Research Program of Shaanxi (Program No. 2023-JC-YB-017), Shaanxi Fundamental Science Research Project for Mathematics and Physics (Grant No. 22JSQ013), “Young Talent Support Plan” of Xi'an Jiaotong University (Grant No.\ WL6J004), the Open Project of State Key Laboratory of Surface Physics (Grant No.\ KF2023\_06), and the Xiaomi Young Talents Program. L. B. thanks the Vannevar Bush Faculty Fellowship Grant No. N00014-20-1C2834 from the Department of Defense and Grant No. DMR-1906383 from the National Science Foundation Q-AMASE-i Program (MonArk NSF Quantum Foundry).\\

\bibliography{Main}

    \newpage
	\textbf{For TOC only}\\
	~~~\\
	~~\\
        \setlength{\belowcaptionskip}{0.1cm}
		\setlength{\abovecaptionskip}{0.1cm}
		\centering
		\includegraphics[scale=0.75]{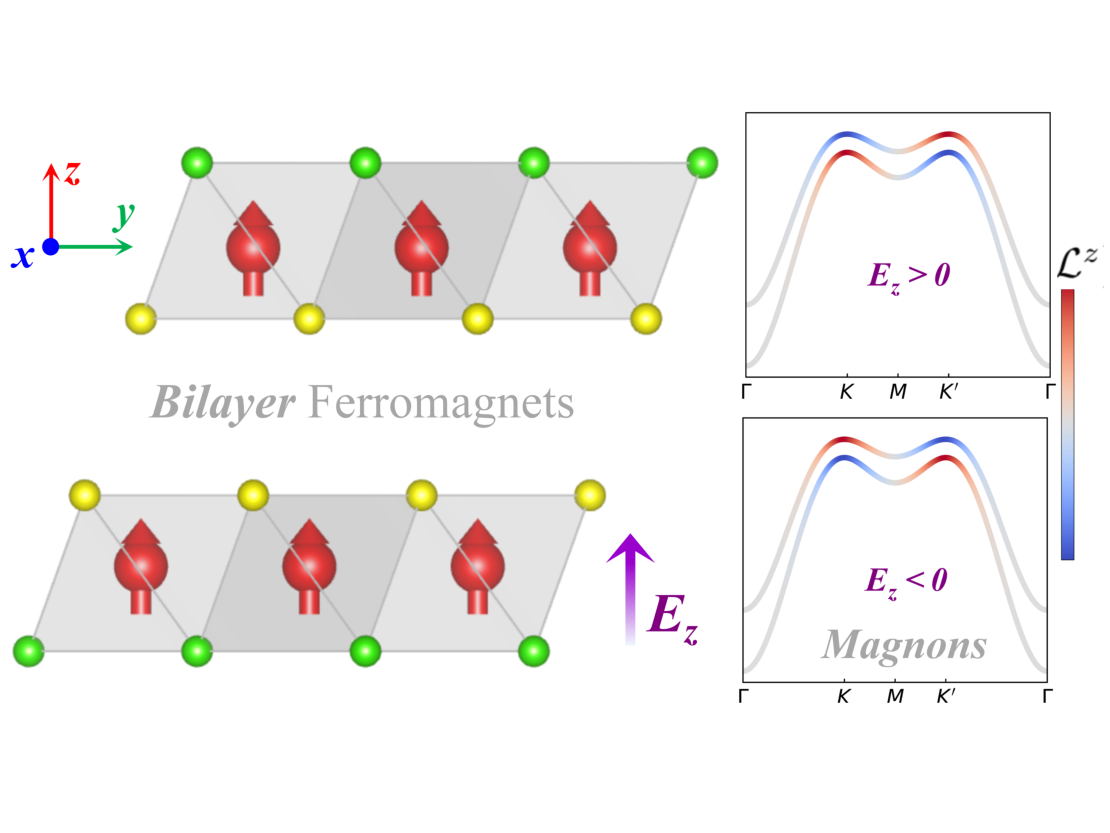}
		\centering
 
\end{document}